\documentstyle[12pt]{article}

\input tcilatex

\begin{document}

\centerline{\large {\bf The boson peak in structural and orientational glasses of simple alcohols:}}  
\centerline{\large {\bf Specific heat at low temperatures}}  \bigskip
\bigskip
\centerline{M. A. Ramos$^{*}$, C. Tal\'{o}n, S. Vieira}  \smallskip
\centerline{\it Laboratorio de Bajas Temperaturas, Depto. de F\'{\i}sica de la Materia Condensada, C-III}
\centerline{\it Instituto ''Nicol\'{a}s Cabrera'', Universidad Aut\'{o}noma de Madrid, Cantoblanco, 28049 Madrid, Spain}
\bigskip
\bigskip
\baselineskip 25pt \centerline{ABSTRACT}

{\small We review in this work specific-heat experiments, that we have conducted on different hydrogen-bonded glasses during last years. Specifically, we have measured the low-temperature specific heat $C_p $ for a set of glassy alcohols: normal and fully-deuterated ethanol, 1- and 2- propanol, and glycerol. Ethanol exhibits a very interesting polymorphism presenting three different solid phases at low temperature: a fully-ordered (monoclinic) crystal, an orientationally-disordered (cubic) crystal or 'orientational glass', and the ordinary structural glass. By measuring and comparing the low-temperature specific heat of the three phases, in the 'boson peak' range 2--10 K as well as in the tunneling-states range below 1K, we are able to provide a quantitative confirmation that ''glassy behavior'' is not an exclusive property of amorphous solids. On the other hand, propanol is the simplest monoalcohol with two different stereoisomers (1- and 2-propanol), what allows us to study directly the influence of the spatial rearrangement of atoms on the universal properties of glasses. We have measured the specific heat of both isomers, finding a noteworthy quantitative difference between them. Finally, low-temperature 
specific-heat data of glassy glycerol have also been obtained. Here we propose a simple method based upon the 
soft-potential model to analyze low-temperature specific-heat measurements, and we use this method for a quantitative comparison of all these data of glassy alcohols and as a stringent test of several universal correlations and scaling laws suggested in the literature. In particular, we find that the interstitialcy model for the boson peak [A. V. Granato, Phys. Rev. Lett. 68 (1992) 974] gives a very good account of the temperature $T_{\rm max}$ at which the maximum in $C_{p} / T^{3}$ occurs.}
\bigskip

--------------------------------------- \\
$^{*}$ Corresponding author. Tel: +34-91 397 5551. Telefax: +34-91 397 3961.
E-mail: miguel.ramos@uam.es

\newpage
\noindent
{\bf 1. Introduction} \bigskip

After several decades of research on the subject, the universal properties exhibited by glasses at low temperatures \cite{zepo,phil} (i.e., their vibrational excitations at low frequencies) continue to be a vivid matter of interest and debate \cite{congresos}. It is well known \cite{elliott,esquinazi} that glasses or amorphous solids have thermal properties (and also dielectric or acoustic ones) very different from those of crystalline solids. Moreover, these properties are very similar among different families of glassy materials irrespective of either the type of chemical bonding or other structural details, hence the name ``universal''. 
At temperatures $T < 1$~K, the specific heat $C_p $ depends approximately linear ($C_p \propto T$) and the thermal conductivity $\kappa$ almost quadratically ($\kappa \propto T^{2}$) on temperature, in 
contrast to the cubic dependences observed in crystals for both properties which is well understood in terms of 
Debye's theory. Quantitatively, the specific heat of nonmetallic glasses is much larger and the thermal conductivity orders of magnitude smaller than those of dielectric crystals. At $T > 1$~K, $C_p $ still deviates from the expected 
$C_{\rm Debye} \propto T^3$ dependence, presenting a broad maximum in $C_{p} / T^{3}$, in the same temperature 
range where the thermal conductivity exhibits a universal {\em plateau}. It is nowadays clear that this universal feature is closely related to an excess in the vibrational density of states $ g ( \nu )$ over the crystalline Debye behavior, leading to a ubiquitous  maximum in $ g ( \nu ) / \nu ^2$ at frequencies $\nu \, \sim 1$~THz which is known as the {\em boson peak}, a dominant feature in the vibrational spectra of glasses very thoroughly observed and studied \cite{congresos} by Raman and inelastic neutron scattering.  

About 30 years ago, Phillips \cite{phil72} and Anderson, Halperin and Varma \cite{AHV} introduced independently 
the now well-known tunneling model (TM), which postulated the ubiquitous existence of atoms or small 
groups of atoms in amorphous solids which can tunnel between two configurations of very similar energy. 
This simple model of tunneling states successfully explained the low-temperature properties of amorphous solids 
\cite{phil}, though only for $T < 1$~K. However, the also rich universal behavior of glasses above 1~K (the broad maximum in $C_{p} / T^{3}$, the corresponding boson peak in vibrational spectra, or the abovementioned plateau in the thermal conductivity) still were unexplained. Among the different approaches proposed since then to understand the general 
behavior of glasses in the whole range of low-frequency excitations, the phenomenological soft-potential 
model (SPM), which can be regarded as an extension of the TM, is one of the best accepted and most often considered. The SPM \cite{kki,ikp} postulates the coexistence in glasses of acoustic phonons (crystalline-like, extended lattice vibrations) with quasilocalized low-frequency ({\em soft}) modes. In the SPM, the potential of these soft modes is assumed to have a uniform stabilizing fourth-order term $W$, an energy scale which is the basic parameter of the model \cite{kki,ikp,bggprs,grbb,gurevich,rb}. In addition, each mode has its own first-order (asymmetry $D_1$) and second-order (restoring force $D_2$) terms, which can be either positive or negative, hence giving rise to a distribution of double-well potentials (tunneling states) and more or less harmonic single-well ones (soft vibrations). 
The parameter $W$ marks the crossover from the tunneling-states region at the lowest 
temperatures to the soft-modes region above it. Indeed, $W$ can be approximately determined either from the minimum $T_{\rm min}$ in $C_{p} / T^{3}$ ($W \simeq \, $1.8--2$ \, k_{\rm B} T_{min}$) or  from the position of the maximum $T_{\kappa, max}$ in a $\kappa/T$ versus $T$ plot:
($W  \simeq \, 1.6 \, k_{\rm B} T_{\kappa, max} $) \cite{rb,rb-esqui}.
 Similarly to the TM, a random distribution of potentials is assumed: $P (D_1 , \, D_2) = P_s $. For a more detailed description of the SPM, the reader is referred to the reviews of Refs. \cite{parshin,rb-esqui}.

In addition to these phenomenological models, other recent approaches \cite{liu,sokolov93,sokolov97,zhu96,zhu98} have focused on suggesting general scalings, correlations or universalities in the low-temperature properties of glasses which could hint at their microscopic origin. 

In order to gain understanding in this issue, we have conducted a series of measurements \cite{prl97,prb98,EtOH2001,POH2001,glycerol,lt22} of specific heat at low temperatures for a special family of glasses, simple alcohols such as ethanol \cite{prl97,prb98,EtOH2001}, propanol \cite{POH2001} and glycerol \cite{glycerol}, which have low glass transition temperatures $T_{\rm g}$ (they are liquid at room temperature) and a molecular, hydrogen-bonded network. In this work, we propose a systematic method to analyze low-temperature specific-heat data, partly based upon the SPM. Then, we employ this method for a critical comparison of the data obtained by us for the alcohols and, finally, we make use of the collected set of parameters as a test of several correlations and scaling laws suggested in the literature.   
\bigskip

\noindent
{\bf 2. Experimental} \bigskip

During last years, we have performed specific-heat measurements on different glassy alcohols, by employing a calorimetric cell especially designed for samples which are liquid at room temperature and that allows to prepare {\em in situ} different solid phases. Firstly, experiments were conducted in a $^4$He cryostat, reaching temperatures down to $\sim $1.7~K. Later, we have used a very similar calorimetric cell within a $^3$He cryostat being able to measure the specific heat to about 0.5~K. 

In particular, we have studied \cite{prl97,prb98,EtOH2001} both normal hydrogenated and fully-deuterated ethanol,
both of which exhibit a rich polymorphism, presenting stable (monoclinic) crystal,
orientational glass (OG, an orientationally-disordered cubic crystal, obtained by quenching a rotationally-disordered plastic phase), and structural glass (amorphous) phases \cite{srini}. 

We have also investigated \cite{POH2001} the behavior of the next substance in the series of monoalcohols, propanol, which is the smallest one which has two different stereoisomers, 1- and 2-propanol, hence allowing us to study directly the effect of the spatial rearrangement of atoms on the low-temperature properties of glasses.

Finally, we have also measured \cite{glycerol} the specific heat and the thermal conductivity of glassy glycerol, probably the most widely studied \cite{wuttke} glass-forming liquid. 

Further details on the experimental setup employed for the heat-capacity measurements, as well as on the experimental procedures followed to obtain and characterize the different solid phases are given in the corresponding references indicated above.
\bigskip

\noindent
{\bf 3. Results} \bigskip

The specific heat of several glassy alcohols (deuterated ethanol in either true glass or OG phases, both isomers of propanol, and glycerol) is shown in Figs. 1 and 2, all of them exhibiting the usual ``glassy'' behavior, with a quasilinear contribution at very low temperature (tunneling states) and a broad maximum in $C_p/T^3$ (boson peak).

First of all, we remark that both structural glass and OG (i.e., a crystal with orientational disorder) phases of ethanol show, qualitatively and even quantitatively, the same glassy features in the low-temperature specific heat \cite{prb98,EtOH2001}, the boson peak in inelastic neutron scattering \cite{prl97}, or at the glass transition \cite{prb98}.
These results provide a {\it quantitative} confirmation of the fact that ``glassy'' behavior is not an exclusive 
property of amorphous solids, which simply lack translational disorder, but a more general characteristic of solids where any kind of disorder is able to {\em soften} the rigid vibrational spectrum of a crystalline lattice and/or to provide additional thermodynamic degrees of freedom which are somehow the basic ingredient of the glass state.  

It is also noteworthy that 2-propanol has a much larger specific heat above 1~K than 1-propanol (and than any other glassy alcohol). However, the reason for this difference is based on its significantly larger Debye contribution (see Fig. 1 and Table 1). This difference between both isomers of propanol also occurs in their crystalline states \cite{POH2001}, which have been recently found to be different indeed, a monoclinic crystalline structure for 1-propanol and a triclinic one for
2-propanol \cite{POH-prl}. Therefore, the influence of the position of the hydroxyl (OH--) on the elastic constants of the hydrogen-bonded network seems to be very relevant. In contrast, the ``excess'' specific heat attributable to tunneling states and quasilocalized vibrations in glasses is much less affected by these changes in the atomic arrangement.

In addition, we have concurrently measured the specific heat and the thermal conductivity of glassy glycerol \cite{glycerol}, and have used those combined data as a more reliable test of the SPM, which has been shown to successfully explain the specific heat and the thermal conductivity in a wide temperature range. In this work, we will only use the specific-heat data for a general comparison with other glassy alcohols.

In Fig. 1, the specific heat of five glasses (four truly amorphous and one disordered crystal) is displayed below $\sim $2.5~K in a typical $C_p/T$ vs $T^2$ plot.
In the simplest version of the TM \cite{phil,phil72,AHV}, the random distribution of tunneling states can be regarded as a constant density of two-level systems (TLS), which neglecting some logarithmic corrections gives a linear contribution to the specific heat $C_p  \, = \, C_{\rm TLS} T $. Taking also into account the Debye contribution due to extended 
long-wavelength vibrations of the amorphous lattice, $C_{\rm Debye} \, = \, C_{\rm D} T^3 $, it is clear that $C_p/T$ vs $T^2$ should plot linearly with an intercept $C_{\rm TLS}$ at $T = 0$.
Although this simple method has been traditionally used to determine $C_{\rm TLS}$ and $C_{\rm D}$, it poses some problems. As Fig. 2 reminds, there is an additional source of specific heat arising from the low-frequency vibrations responsible for the boson peak and the maximum in $C_{p} / T^{3}$. This contribution is not completely negligible below      1~K, that brings as a consequence that many linear fits $ C_p  \, = \, C_{1} T \, + C_{3} \, T^3 $ found in the literature unavoidably provide cubic coefficients clearly exceeding the true Debye one, obtained from elastic measurements in those cases where they are available, i.e., $C_3 > C_{\rm D}$. Indeed, $C_{1}$ and especially $C_{3}$ depend on the chosen range for the linear fit. In order to solve these inconveniences, we propose a systematic method to analyze low-temperature specific-heat measurements. The basic point is to realize that the difficulties originate from the lower-energy side of the additional vibrations responsible for the boson peak. These have been well accounted for by the SPM
\cite{parshin,rb-esqui} as harmonic soft modes giving rise to a specific heat below the maximum
$ C_p \, = C_{\rm sm} T^5 = \frac{2 \pi^6}{21} P_s k_{\rm B} \left(\frac{k_{\rm B}T}{W}\right)^{5} $.
Therefore, a quadratic polynomial fit in  $C_p/T$ vs $T^2$ seems the most appropriate solution.
The question however remains how to decide the temperature range to fit the data, with physical meaning.
It is clear that the distribution of soft modes and correspondingly the specific heat can not grow $C_p \propto T^5$ unlimitedly.
Gil {\em et al.} \cite{grbb} proposed a gaussian distribution in the asymmetry of the soft potentials, which without any further fitting parameter allowed them to account for the specific heat, thermal conductivity and vibrational density of states $g (\nu ) / \nu^2$ in the whole relevant range, including the ``boson peak''. At least for practical reasons, let us assume that distribution function. It is easy to find that the simple $C_{\rm sm} T^5$ approximation starts deviating $\sim$~5\% from the $C_p (T)$ curve accounting for experimental data approximately at $T > 0.75 W \, \simeq  \, \frac{3}{2} T_{\rm min}$. Therefore, we suggest to fit specific-heat data in a $C_p/T$ vs $T^2$ representation by using a quadratic polynomial $C_p  \, = \, C_{\rm TLS} T \, + \, C_{\rm D} T^3 + C_{\rm sm} T^5$ in the temperature range 
$ 0 < T < \frac{3}{2} T_{\rm min}$. The results from these fits for the studied glassy alcohols are shown in Table 1.
As a proof of consistency, we want to mention that the obtained $C_{\rm D}$ for glycerol agrees better than 1\% (hence within the experimental error) with the Debye coefficient estimated \cite{glycerol} from elastic measurements. Unfortunately, this is the only glass from those studied here for which elastic data are available. Nevertheless, we believe that the method proposed is a reasonable alternative to determine the Debye coefficient of a glass from low-temperature specific-heat measurements, especially when elastic data are lacking. The so-obtained values of $C_{\rm D}$ and $\theta_{\rm D}$ have been used in Fig. 2 to scale $C_p$ data for the reasons discussed below. 

For sake of completeness, we also show in Table 1 the SPM parameters $W$ (determined from $W \simeq \, 1.8 \, k_{\rm B} T_{min}$ \cite{grbb,rb-esqui}, with the exception of glycerol which has a flat minimum and has been better determined from available thermal-conductivity data \cite{glycerol}) and $P_s$ (determined from the given expression for $C_{\rm sm}$). It is to be stressed that all studied glasses present comparable values of $P_s$ (basically, the distribution density of quasilocalized excitations, either tunneling states or soft modes), not existing any significant difference between amorphous glasses and orientational glasses (OG).
\bigskip

\noindent
{\bf 4. Discussion} \bigskip

In this section, we will make use of the compiled set of data for the studied glassy alcohols taken as a model system to critically discuss several correlations or scaling laws which have been suggested in the literature to be universal for glasses or amorphous solids.

First, we will address very briefly the scaling proposed by Liu and L\"{o}hneysen \cite{liu}. They suggested a general correlation between the mechanisms leading to the $C_{p}/T^{3}$ maximum in crystalline and amorphous solids. 
They plotted the height $P_{c}$ of the maximum in $C_{p}/T^{3}$ (i.e., $(C_p/T^3)_{\rm max}$ in our notation) versus its position $ T_{\rm max}$ for a wide set of materials, mainly amorphous polymers and metals, as well as typical network glasses, together with their corresponding crystalline solids. They found an approximate general correlation
$P_{c} \propto T_{max}^{-1.6}$, somehow indicating a close relation between the $C_{p}/T^{3}$ maxima in glasses and crystals. However, we have included in such a graph (see Fig. 2 of Ref. \cite{lt22}) specific-heat data for molecular glasses and crystals (both our hydrogen-bonded materials and van der Waals glasses from Lindqvist {\em et al.} \cite{lindqvist}), finding that these molecular solids deviate systematically about a factor 5 from the general scaling proposed by those authors. 
Moreover, for the different phases of H- and D-ethanol we have plotted  $(C_p / T^3 ) / P_c $ vs $T/T_{\rm max}$, as also suggested by Liu and von L\"ohneysen \cite{liu}. We found (see Fig. 11 of Ref. \cite{prb98}) that data for the disordered solids (either glass or OG) superimpose in the whole temperature 
region but, in contrast, the curves for the ordered crystals showed a far narrower shape. This fact points out again to the 
different nature of the low-energy vibrational spectra of glasses and crystalline solids.

Other authors have suggested that the low-energy excitations of glasses may be correlated with the {\em fragility} of the glass-forming liquid, a parameter proposed by Angell \cite{angell,NIST} to characterize how fragile or strong is a supercooled liquid to resist the structural degradation induced by temperature, and which is usually measured by the deviation of the shear viscosity from an Arrhenius law. Zhu \cite{zhu96} has suggested a general correlation between the density of tunneling states and the fragility, presenting data for a variety of glasses. More fragile glasses would have a larger number of minima on the potential-energy hypersurface, hence explaining a higher density of tunneling states and a higher value of $C_{\rm TLS}$. On the other hand, Sokolov {\em et al.} \cite{sokolov97} have correlated the fragility of the system with the excess of vibrational excitations normalized to the Debye level. Specifically, they showed that the ratio  
$(C_{exc}/C_{\rm D})_{\rm max} \, = \, (C_{p}/C_{\rm D} \, - \, 1 )_{\rm max} $ decreases with increasing fragility. More recently, Zhu and Chen \cite{zhu96} have argued against the universality of such a correlation. 

Unfortunately, accurate fragility values using the same criteria are not available for the four glass-forming liquids studied in this work. Nevertheless, it can be seen from viscosity data \cite{angell,NIST} that all alcohols possess an intermediate degree of fragility. Liquid ethanol seems to be slightly more fragile than glycerol, and 1-propanol is less fragile. There are no published data for 2-propanol to our knowledge. In this framework, the large difference in $ C_{\rm TLS}$ between ethanol and glycerol (almost one order of magnitude, see Table 1) for two systems of very similar fragility claims against the correlation proposed by Zhu \cite{zhu96}. Moreover, 1-propanol has a contribution to $C_p$ from tunneling states almost a factor of 3 higher than that of glycerol, being less fragile. The correlation with $(C_{exc}/C_{\rm D})_{\rm max}$ is more difficult to ascertain. The obtained values are all around $(C_{exc}/C_{\rm D})_{\rm max} \, \sim  \, 0.5$ as expected for intermediate fragilities, but the smaller deviations do not show up the expected trend.
Neither suggest normalized $C_{p} / C_{\rm D} $ curves shown in Fig. 2 any kind of universal behavior. 
 Furthermore, the very similar values of $ C_{\rm TLS}$ and $(C_{exc}/C_{\rm D})_{\rm max}$ for either true glasses (quenching a supercooled liquid) or orientational glasses (quenching a plastic crystal) of ethanol cast doubts about the relevance of the fragility of the glass-forming liquid to understand the low-temperature properties of glasses.
  
Finally, we would like to consider the interstitialcy model proposed by Granato \cite{granato1,granato2}. According to this model, liquids can be considered as crystals containing a few interstitials in thermal equilibrium, which become frozen in the glassy state. Self-interstitial resonance modes would be the physical realization of the soft modes and tunneling states of the SPM, giving rise to the boson peak. For the simplest approximation, taken a single frequency for the resonance modes, the maximum in $C_{p} / T^{3}$ is predicted \cite{granato2} to appear at a temperature $T_{\rm max} \, \sim \, \theta_{\rm D} / 35$. Several authors have tested this relation for different collections of experimental data, finding empirically $T_{\rm max} \approx \theta_{\rm D} / 40$ \cite{carini} or $T_{\rm max} \approx \theta_{\rm D} / 38$ \cite{zhu98}. It can be seen in Table 1 that this correlation works very well for all glassy alcohols studied by us. On the contrary, the maxima of the corresponding crystalline states \cite{prb98,POH2001,glycerol} take place at higher reduced temperatures ($T_{\rm max} / \theta_{\rm D} \, \sim \, 1/25 $ for these alcohols), as it seems always be the case (see Fig. 3.3 in Ref. \cite{phil}). 
However, we have tested this correlation between $T_{\rm max}$ and $\theta_{\rm D} / 35$ from published data  \cite{eloy} for another model system: vitreous B$_2$O$_3$ submitted to different thermal treatments producing significant changes in mass density, elastic constants, etc. \cite{b2o3}. Variations of up to 17\% in $T_{\rm max}$ and $\theta_{\rm D}$ were achieved with this method. We find that, although $(C_{exc}/C_{\rm D})_{\rm max}$ ranges uncorrelatedly from 1.18 to 1.70 for the seven different B$_2$O$_3$ glasses, the ratio between $T_{\rm max} $ and $ \theta_{\rm D} $
remains constant for all of them: $ \theta_{\rm D} / T_{\rm max} \, = \, 51 \pm 1 $. Let us notice that for other oxide glass as SiO$_2$, $ \theta_{\rm D} / T_{\rm max} \, = \, 46$ \cite{zhu98}. 
So, it may well be that the exact $T_{\rm max} /\theta_{\rm D}$ ratio could somewhat depend on the kind of glass, what is not in conflict with the interstitialcy model \cite{granato2}, but for a given ``system'', fixing some secondary parameters, the boson peak and Debye temperatures seems to be clearly correlated. 
\bigskip

\noindent
{\bf 5. Conclusions} \bigskip

In summary, we have reviewed and comparatively discussed specific-heat experiments, that we have conducted on different hydrogen-bonded glasses during last years: normal and fully-deuterated ethanol, 1- and 2- propanol, and glycerol. 
We have proposed a systematic method partly based upon the soft-potential model to analyze these low-temperature specific-heat measurements, and to test several universal correlations and scaling laws suggested in the literature. In particular, we have found that the correlation  between the temperature $T_{\rm max}$ at which the maximum in $C_{p} / T^{3}$ occurs and the Debye temperature $\theta_{\rm D}$ proposed by the interstitialcy model for the boson peak is very well fulfilled. 
\bigskip 

\noindent
{\bf Acknowledgements} \bigskip

This work was supported by MCyT (Spain) within project BFM2000-0035-C02-01.

      \newpage

\noindent
{\bf FIGURE CAPTIONS} \bigskip

FIGURE 1: Low-temperature specific heat plotted as $C_p/T$ vs $T^2$ for several glassy alcohols. Solid lines are fits to quadratic polynomials (see text). \\

FIGURE 2: Low-temperature specific heat of the same glasses in Fig. 1, scaled to the cubic Debye contribution $C_{\rm Debye}$, as a function of temperature normalized to $\theta_{\rm D}$. The arrow indicates the position of the maximum at $T = \theta_{\rm D}/35$ predicted within the interstitialcy model. \\

\newpage
\noindent
{\bf TABLES} \bigskip

\begin{table}[h]
\begin{center}
\begin{tabular}{lcccccccccccc}
&
\multicolumn{2}{c}{H-ethanol}&
\multicolumn{2}{c}{D-ethanol} &
\multicolumn{1}{c}{1-propanol} &
\multicolumn{1}{c}{2-propanol} &
\multicolumn{1}{c}{glycerol} \\
&
glass$^*$ & OG & 
glass & OG & 
glass & 
glass & 
glass\\
\hline
$P_{\rm mol}$ (g/mol) &
\multicolumn{2}{c}{46.1} &
\multicolumn{2}{c}{52.1} &
\multicolumn{1}{c}{60.1} &
\multicolumn{1}{c}{60.1} &
\multicolumn{1}{c}{92.1} \\
$T_{\rm g}$ (K) &
95 & 95 & 
95 & 95 & 
98 & 
115 & 
185  \\
\hline
$T_{\rm min}$ (K) &
2.3 & 2.6 & 
2.1 & 2.3 & 
1.8 & 
1.6 & 
2.0  \\
$T_{\rm max}$ (K) &
6.1 & 6.8 & 
6.1 & 6.4 & 
6.7 & 
5.0 & 
8.7  \\
$(C_p/T^3)_{\rm max}$ &
2.4 & 2.2 &
2.8 & 2.6 &
2.7 & 
3.6 &
1.4  \\
{\small (mJ/mol·K$^4$)} & & & & & & \\
\hline
$ C_{\rm TLS}$ (mJ/mol·K$^2$)&
1.2 & 1.27 &
1.05 & 1.13 & 
0.424 & 
0.516 & 
0.157  \\
$C_{\rm D}$ (mJ/mol·K$^4$)&
1.55 & 1.45 & 
1.80 & 1.72 & 
1.77 & 
2.54 & 
0.855  \\
$C_{\rm sm}$ (mJ/mol·K$^6$)&
0.0432 & 0.0288 & 
0.0572 & 0.0419 & 
0.0367 & 
0.0845 & 
0.0139 \\
\hline
$W$ (K) &
4.1 & 4.7 & 
3.8 & 4.1 & 
3.3 &
2.9 & 
4.3$^\dagger$  \\
$P_s$ (mol$^{-1}$)&
4.0$\times10^{19}$ & 5.2$\times10^{19}$ & 
3.6$\times10^{19}$ & 3.8$\times10^{19}$ & 
1.1$\times10^{19}$ & 
1.4$\times10^{19}$ & 
1.6$\times10^{19}$  \\
\hline
$(C_{exc}/C_{\rm D})_{\rm max}$ &
0.55 & 0.52 & 
0.56 & 0.51 &
0.53 & 
0.42 & 
0.64  \\
$\theta_{\rm D}$ (K)&
224 & 229 & 
213 & 217 & 
236 & 
209 & 
317  \\
$\theta_{\rm D} / T_{\rm max} $&
37 & 34 & 
35 & 34 & 
35 & 
42 & 
36  \\
\hline
\end{tabular}
\bigskip
\caption{}{Measured data and fit parameters obtained for several studied glassy alcohols (see text).
$^*$ Data for the glass phase of H-ethanol were taken in a $^4$He-cryostat, only down to 1.7 K.
$^\dagger$ The value of $W$ for glycerol has been determined from thermal conductivity data \cite{glycerol}.}
\end{center}
\end{table}
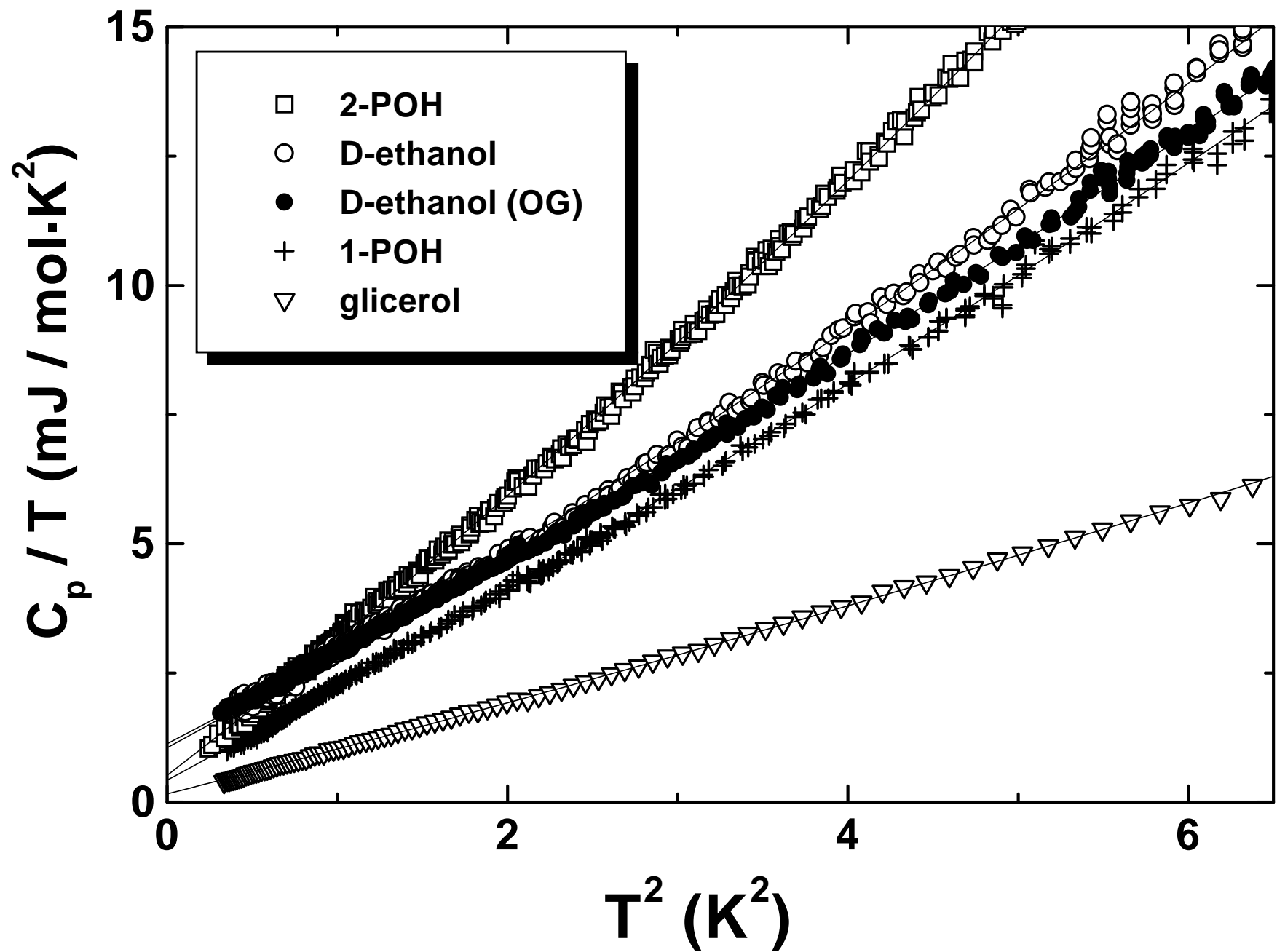
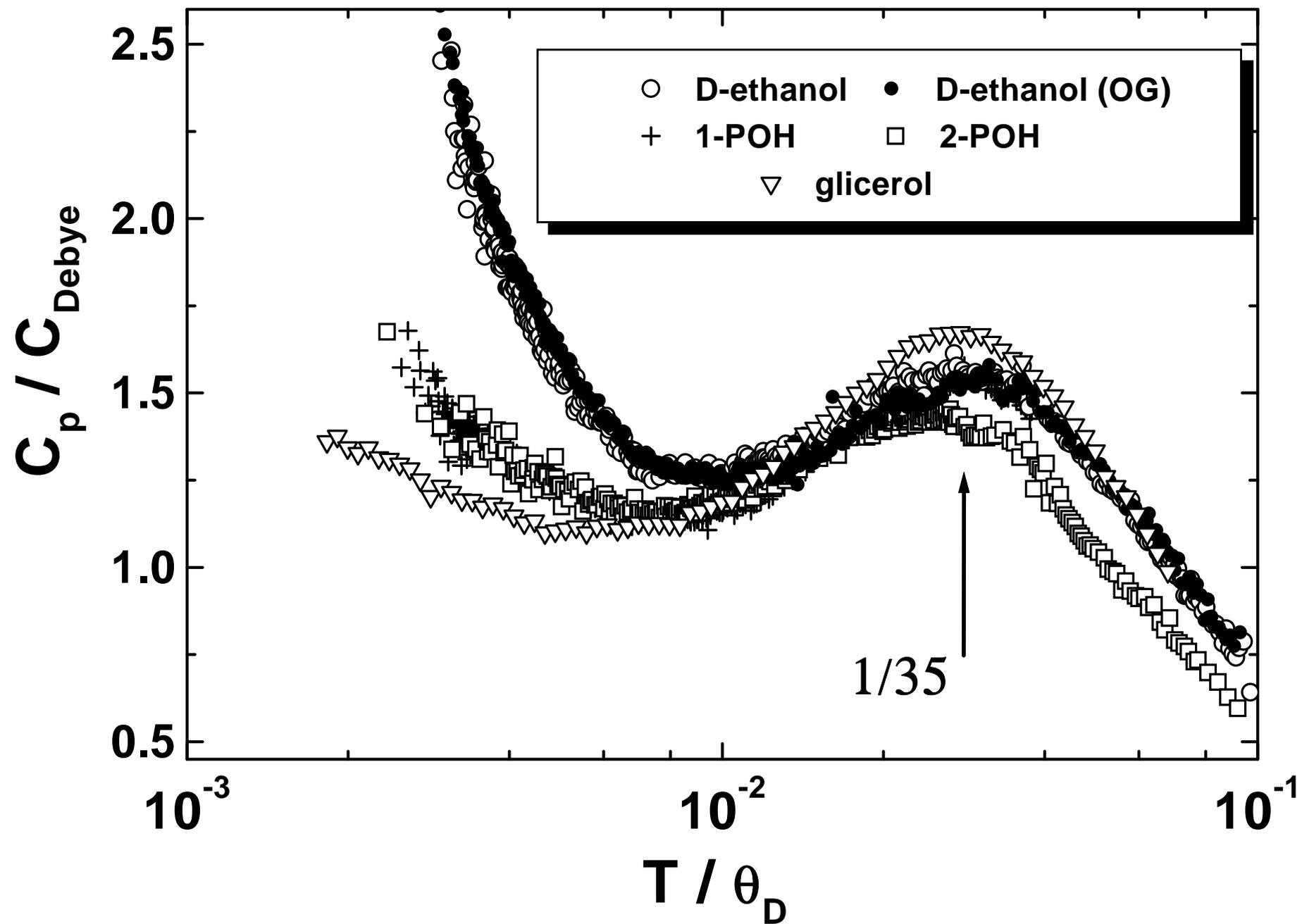
\end{document}